\documentclass[apj]{emulateapj}

\usepackage{natbib}

\usepackage[hang,raggedright,tight]{subfigure}
\usepackage{longtable}
\usepackage[figuresright]{rotating}
\usepackage{float}
\usepackage[toc,page]{appendix}

\def\gsim{\;\rlap{\lower 2.5pt
\hbox{$\sim$}}\raise 1.5pt\hbox{$>$}\;}
\def\lsim{\;\rlap{\lower 2.5pt
\hbox{$\sim$}}\raise 1.5pt\hbox{$<$}\;}
\def\ie{{\it i.e.}}
\def\eg{{\it e.g. }}

\usepackage{xcolor}

\newcommand{\be}{\begin{equation}\rm}
\newcommand{\ee}{\end{equation}}
\usepackage{amsmath}

\shorttitle{Model-Independent Halo Mass Function Measurement}
\shortauthors{Dong et al.}

\begin{document}

\title{Towards a Model-Independent Measurement of the Halo Mass Function with Observables}

\author{Fuyu Dong$^1$, Jun Zhang$^{1,2*}$, Xiaohu Yang$^{1,2,3}$, Jiajun Zhang$^1$, Wentao Luo$^4$}
\affil{
$^1$Department of Astronomy, Shanghai Jiao Tong University, Shanghai 200240, China \\
$^2$Shanghai Key Laboratory for Particle Physics and Cosmology, Shanghai 200240, China \\
$^3$Tsung-Dao Lee Institute, Shanghai Jiao Tong University, Shanghai 200240, China\\
$^4$Institute for the Physics and Mathematics of the Universe (Kavli IPMU, WPI), UTIAS,\\ Tokyo Institutes for Advanced Study, University of Tokyo, Chiba, 277-8583, Japan
}

\email{*betajzhang@sjtu.edu.cn}

\begin{abstract}
In the CDM paradigm, the halo mass function is a sensitive probe of the cosmic structure. In observations, halo mass is typically estimated from its relation with other observables. The resulting halo mass function is subject to systematic bias, such as the Eddington bias, due to the scatter or uncertainty in the observable - mass relation. Exact correction for the bias is not easy, as predictions for the observables are typically model-dependent in simulations. In this paper, we point out an interesting feature in the halo mass function of the concordence $\Lambda$CDM model: the total halo mass within each evenly-spaced logarithmic mass bin is approximately the same over a large mass range. We show that this property allows us to construct an almost bias-free halo mass function using only an observable (as a halo mass estimator) and stacked weak lensing measurements as long as the scatter between the true halo mass and the observable-inferred mass has a stable form in logarithmic units. The method is not sensitive to the form of the mass-observable relation. We test the idea using cosmological simulations, and show that the method performs very well for realistic observables.

\end{abstract}

\section{INTRODUCTION}
\label{introduction}
In the CDM cosmological models, dark matter halos grow hierarchically from small perturbations in the initial density field \citep{1993MNRAS.262..627L}. The abundance of halos as a function of the halo mass is an important probe for cosmology \citep{1974ApJ...193..437P,1993MNRAS.262.1023W}. In observations, halo mass can be estimated with certain observables, such as the total luminosities of the member galaxies \citep{2004MNRAS.353..189V}, the number of the member galaxies\citep{2009ApJ...703.2217S,2012ApJ...749...56B,2008JCAP...08..006M}, the X-ray emission from hot gas \citep{2009A&A...498..361P}, the gravitational lensing signals \citep{2001PhR...340..291B,2005astro.ph..9252S}, the Sunyaev-Zel\'dovich effect \citep{1972CoASP...4..173S}, the galaxy velocity dispersion \citep{2006A&A...456...23B,2013MNRAS.430.2638M}, etc.. In terms of constraining cosmology, recent efforts have mainly focused on combining the halo abundance with their spatial clustering information \citep{2005PhRvD..72d3006L}, preferably with a comprehensive modelling of the observation with a large mock sample \citep{1998ApJ...494....1J, 2002ApJ...575..587B, 2003MNRAS.339.1057Y,2005ApJ...633..791Z, 2007MNRAS.376..841V, 2011ApJ...736...59Z,2011ApJ...738...45L, 2012ApJ...752...41Y, 2015ApJ...799..130R,2016MNRAS.457.4360Z, 2016MNRAS.459.3040G, 2017MNRAS.470..651R,2018ApJ...858...30G,2018MNRAS.474.5500O}. 

Nevertheless, a direct and accurate measurement of the halo mass function in observation is still not easy. This is due to several reasons: 1) predictions of most observables in simulations are currently model-dependent; 2) the mass - observable relation typically contains a significant scatter, causing the so called Eddington bias \cite{1913MNRAS..73..359E}; 3) the definition of halo mass in simulation has subtle differences for different types of observables; 4) accuracy of weak lensing mass measurement on individual halos is limited by the number density of background source images and the uncertainties of the foreground halo shapes; 5) the completeness of the halo catalog from observation needs to be known very well to form a reasonable comparison with simulation predictions. For the halo mass function to become a useful probe for precision cosmology, the above problems need to be addressed or avoided, ideally in a model-independent way. This is our motivation. We propose a solution that has the potential of avoiding most, if not all, of the above problems in the measurement of the halo mass function. 

Weak lensing is the most direct way of measuring halo mass. So far, lensing is commonly used for calibrating the mass - observable relation through the modeling approach. By grouping the foreground sources according to a particular observable, \eg, luminosity, stellar mass, etc. \citep{2017MNRAS.467.3024L,2017MNRAS.466.3103S,2018ApJ...854..120M}, their stacked background shear distribution are used to constrain the model parameters in the mass - observable relation. Here, we point out that {\it the lensing-reconstructed mean mass of halos binned by sorting a certain observable can sometimes be directly interpreted as an accurate estimation of the halo mass function, albeit in a slightly unconventional form: the halo mass as a function of the mass order/rank within a certain cosmic volume}. This forms an almost model-independent way of estimating the halo mass function, as we show in the rest of the paper. 

We call the new form of the halo mass function the Ranked Halo Mass Function (RHMF). It can almost be regarded as the traditional halo mass function rotated by 90 degrees, except that the halo number density is replaced by the halo mass rank, which is well defined only for a finite volume. Measurement of the RHMF is straightforward in simulation, but not so in observation, as the masses of individual halos are not accurately known. To our surprise, we find that under certain conditions (which is approximately satisfied in real cases), if the mass-ranks of halos are generated by sorting a certain type of observables instead, the resulting mean halo mass in each bin of a given mass-rank range is very close to the true one! This is true despite the fact that halos are not actually correctly ordered by their true masses in this case (due to the dispersion in the mass - observable relation). The reasons for the above phenomenon, as well as the conditions required on the observable, are given in \S\ref{method}. We further show that the mean halo mass in each bin can be well recovered using the stacked weak lensing signals on the background, ultimately leading to an almost unbiased estimate of the RHMF. This is shown in \S\ref{simulation} with simulations and realistic observables. A brief conclusion and discussions are given in \S\ref{discussion}. \\



\begin{figure}
    \centering
     \subfigure{
    \includegraphics[width=0.49\textwidth, clip]{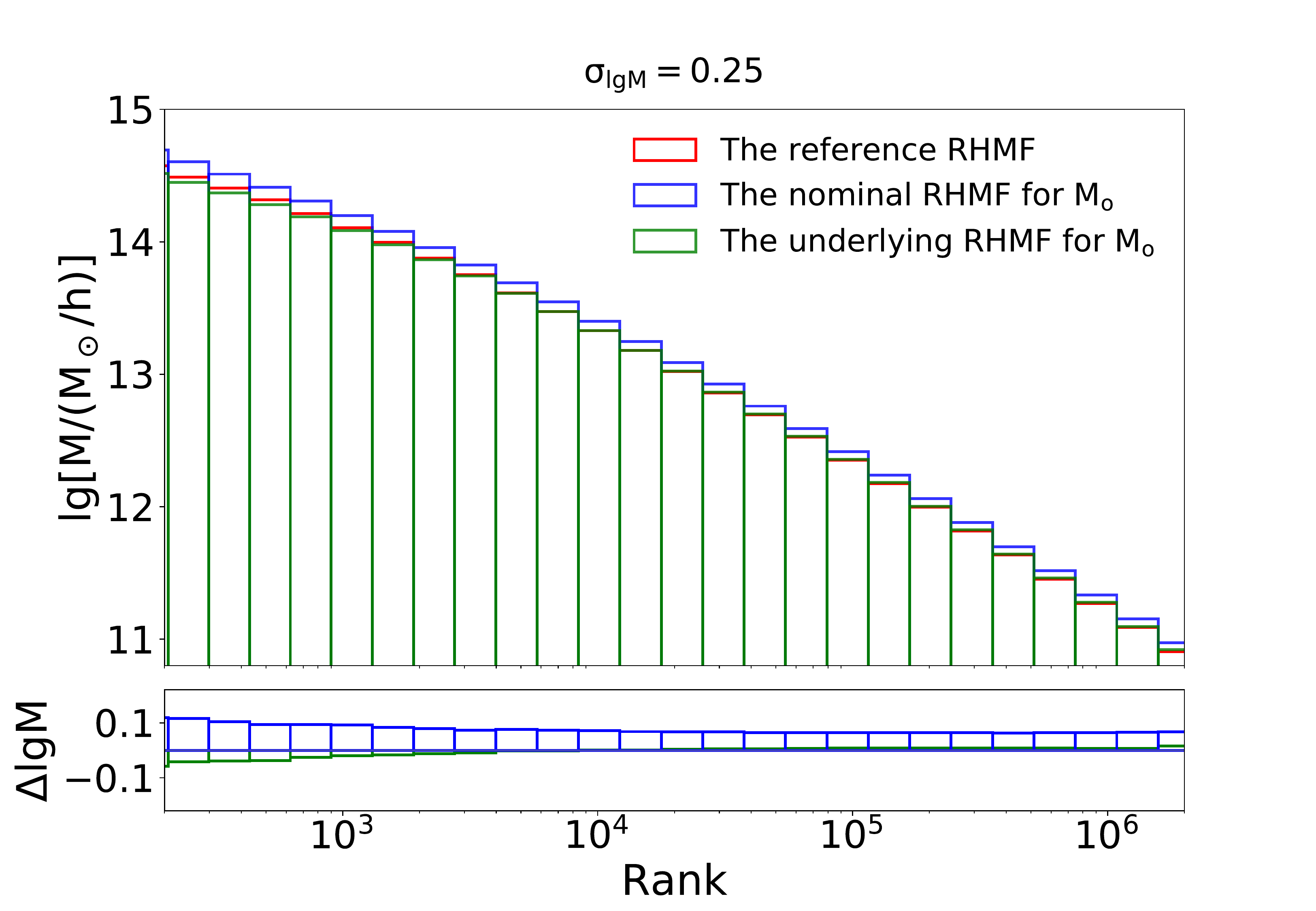}}
    \caption{ \label{fig-hmf-rand} 
    Results for the Ranked Halo Mass Function (RHMF). The red histogram in the upper panel shows the reference RHMF completely generated with the true halo
    masses. The blue histogram shows the nominal RHMF using only the mass $M_{o}$ that is converted from the observable. The green histogram is the underlying RHMF yielded 
    by the true halo mass distribution for bins that are formed by sorting the $M_{o}$. The lower panel shows the difference between the blue/green histograms and the red one.}
\end{figure}

\section{Sorting Halos With Observables}
\label{method}

To illustrate our idea, we generate a set of halo masses using the Sheth - Tormen formula (S-T) of the $\Lambda$CDM universe \citep{2001MNRAS.323....1S}, with $\Omega_c=0.223$, $\sigma_8=0.85$, and $\Omega_b=0.045$. To mimic the halo mass $M_o$ derived from an observable (through certain mass-observable relation), we add a lognormal error to the true halo mass $M_t$ for each halo, with $\sigma_{\rm lgM}=0.25$. In Fig.\ref{fig-hmf-rand}, the red histogram in the upper panel shows the reference RHMF for the $M_t$, \ie, $M_t$ is used for both ranking the halos and calculating the mean halo mass in each bin. It should be regarded as the theoretical prediction, as it is what one can easily derive from analytical formulae or N-body simulations for any given cosmological model. It is interesting to show the blue histogram for a comparison, which uses $M_ o$ for both ranking and calculation of the mean. The result is what we call the nominal RHMF for the $M_o$. The blue histogram can be directly derived from observation once a mass-observable relation is given, as it completely relies on $M_o$. The difference between the blue and red histograms is a result of the Eddington bias. Their ratios are shown in the lower panel of Fig.\ref{fig-hmf-rand}.  

An interesting alternative, which is indeed the main point of this paper, is to consider plotting the mean of the {\it true} halo mass in bins that are formed by sorting $M_o$. The resulting RHMF is shown in the upper panel of Fig.\ref{fig-hmf-rand} with the green histogram. It is called the underlying RHMF for the $M_o$. Its ratio to the red histogram, \ie, the reference RHMF, is shown in the lower panel of the same figure. It is important to note that the green histogram yields almost no systematic biases in the RHMF within a large mass range. This is a very useful feature of the $\Lambda$CDM model, as we show next. We suggest that people who are not interested in this proof can directly move to the paragraph below  eq.(\ref{eqinte}).

For our purpose, it is easier to work with halo masses in logarithmic units. We therefore define $\mathrm{ln}M_t$ as $m_t$, and $\mathrm{ln}M_o$ as $m_o$. For a halo of mass $m_t$, we define $P(m_o\vert m_t)dm_o$ as the probability that the observed mass is in the range of $m_o$ and $m_o+dm_o$. Note that the observed mass here simply refers to the halo mass that is converted from an observable through a mass-observable relation. Let us also define the true halo mass function as $N_t(m_t)dm_t$, \ie, the number of halos with true masses between $m_t$ and $m_t+dm_t$ for a given cosmic volume. Similarly, we define the number of halos with observed masses being in the range of $m_o$ and $m_o+dm_o$ as $N_o(m_o)dm_o$. $N_o(m_o)$ and $N_t(m_t)$ have the following relation:
\begin{equation}
\label{No}
N_o(m_o)=\int_{-\infty}^{\infty}dm_tN_t(m_t)P(m_o\vert m_t)
\end{equation}

The mean true halo mass $\overline{M}_t$ (not in logarithmic unit) that the observed $m_o$ corresponds to is:
\begin{equation}
\label{mbarbar}
\overline{M}_t(m_o)=\int_{-\infty}^{\infty}dm_t\exp(m_t)P'(m_t\vert m_o)
\end{equation}
where $P'(m_t\vert m_o)$ is the probability that $m_o$ corresponds to the true halo mass $m_t$. $P'$ and $P$ are related through:
\begin{equation}
\label{PP}
P'(m_t\vert m_o)N_o(m_o)=P(m_o\vert m_t)N_t(m_t)
\end{equation}
Combining eqs.(\ref{No},\ref{mbarbar},\&\ref{PP}), we get:
\begin{equation}
\label{mbar}
\overline{M}_t(m_o)=\frac{\int_{-\infty}^{\infty}dm\exp(m)P(m_o\vert m)N_t(m)}{\int_{-\infty}^{\infty}dmP(m_o\vert m)N_t(m)}
\end{equation}

On the other hand, for a halo of observed mass $m_o$, the rank $RK_o$ from sorting the observed mass is given by:
\begin{equation}
\label{RK1}
RK_o(m_o)=\int_{m_o}^{\infty}dmN_o(m)
\end{equation}
Assuming the rank $RK_o$ in the true halo mass function corresponds to halo mass $\widetilde{M}_t$ (not in logarithmic unit), we have:
\begin{equation}
\label{RK2}
RK_o(m_o)=\int_{\mathrm{ln}\widetilde{M}_t}^{\infty}dmN_t(m)
\end{equation}   
Eqs.(\ref{No},\ref{RK1},\&\ref{RK2}) lead to:
\begin{equation}
\label{mtilde}
\int_{\mathrm{ln}\widetilde{M}_t}^{\infty}dmN_t(m)=\int_{m_o}^{\infty}dm\int_{-\infty}^{\infty}dm'N_t(m')P(m\vert m')
\end{equation}

Our key point is to find out whether $\widetilde{M}_t$, \ie, the mass of the bin around the rank of $RK_o$ for the true RHMF, can be well approximated by the $\overline{M}_t$ from sorting an observable (the green histograms in Fig.\ref{fig-hmf-rand}). For this purpose, let us assume that the true halo mass function is an exponential function: $N_t(m)\propto \exp(\alpha m)$, with $\alpha$ being a negative number. This form corresponds to a power law halo mass function in linear mass unit, where $\alpha-1$ is the power index. Eq.(\ref{mtilde}) then becomes:
\begin{equation}
\label{mtilde2}
-\frac{1}{\alpha}\widetilde{M}_t^{\alpha}=\int_{m_o}^{\infty}dm\int_{-\infty}^{\infty}dm'\mathrm{e}^{\alpha m'}P(m\vert m')
\end{equation} 
and eq.(\ref{mbar}) becomes:
\begin{equation}
\label{mbar2}
\overline{M}_t(m_o)=\frac{\int_{-\infty}^{\infty}dm P(m_o\vert m)\mathrm{e}^{(1+\alpha) m}}{\int_{-\infty}^{\infty}dmP(m_o\vert m)\mathrm{e}^{\alpha m}}
\end{equation}

Let us further suppose that the function $P(m_o\vert m_t)$ can be written as $f(m_t-m_o)$, \ie, the probability density function only depends on the ratio of the masses in linear units. We can then transform eq.(\ref{mtilde2}) as:
\begin{eqnarray}
\label{mtilde3}
-\frac{1}{\alpha}\widetilde{M}_t^{\alpha}&=&\int_{m_o}^{\infty}dm\int_{-\infty}^{\infty}dm'\mathrm{e}^{\alpha m'}f(m'-m)\\ \nonumber
&=&\int_{m_o}^{\infty}dm\mathrm{e}^{\alpha m}\int_{-\infty}^{\infty}dm''\mathrm{e}^{\alpha m''}f(m'')\\ \nonumber
&=&-\frac{1}{\alpha}\mathrm{e}^{\alpha m_o}\int_{-\infty}^{\infty}dm''\mathrm{e}^{\alpha m''}f(m'')
\end{eqnarray} 
As a result, we get:
\begin{eqnarray}
\label{mtilde4}
\widetilde{M}_t=\mathrm{e}^{m_o}\left\{\int_{-\infty}^{\infty}dm''\mathrm{e}^{\alpha m''}f(m'')\right\}^{\frac{1}{\alpha}}
\end{eqnarray}

Similarly, eq.(\ref{mbar2}) can be rewritten as:
\begin{eqnarray}
\label{mbar3}
\overline{M}_t(m_o)&=&\frac{\int_{-\infty}^{\infty}dm f(m-m_o)\mathrm{e}^{(1+\alpha) m}}{\int_{-\infty}^{\infty}dmf(m-m_o)\mathrm{e}^{\alpha m}}\\ \nonumber
&=&\mathrm{e}^{m_o}\cdot\frac{\int_{-\infty}^{\infty}dm' f(m')\mathrm{e}^{(1+\alpha) m'}}{\int_{-\infty}^{\infty}dm'f(m')\mathrm{e}^{\alpha m'}}
\end{eqnarray}

It is easy to see that when $\alpha=-1$, an interesting phenomenon emerges: $\overline{M}_t=\widetilde{M}_t$. Note that the following normalization condition is used:
\begin{equation}
\label{eqinte}
\int_{-\infty}^{\infty}dm_oP(m_o\vert m_t)=1
\end{equation}
which leads to $\int_{-\infty}^{\infty}dm f(m)=1$. The above calculation provides a reason for the agreement between the green and red histograms in Fig.\ref{fig-hmf-rand}. We see that the conclusion relies on two conditions: 

1) $N_t(m)\propto \exp(-m)$, \ie, the total halo mass within each evenly-spaced logarithmic mass bin is the same;

2) $P(m_o\vert m_t)$ is a function of $m_t-m_o$ only, \ie, the dispersion between the observable-inferred halo mass and the true halo mass in logarithmic units is independent of the true halo mass.


\begin{figure}
    \centering
    \subfigure{
    \includegraphics[width=0.5\textwidth, clip]{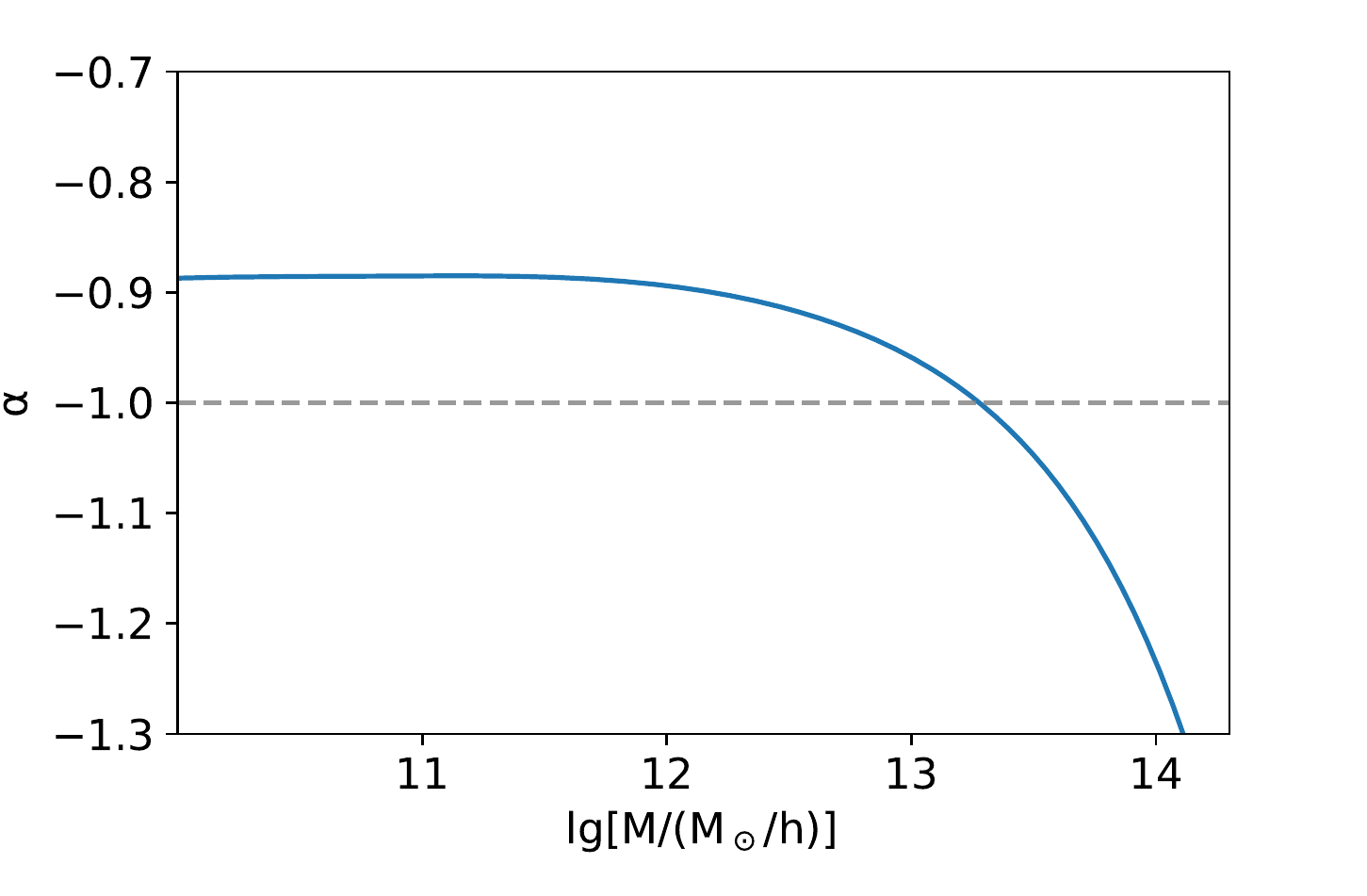}}
    \caption{ The power law index $\alpha$ of the halo mass function ($dn/dlnM\sim M^\alpha$) as a function of the halo mass in the LCDM model.}
    \label{fig-alpha}
\end{figure}


For the $\Lambda$CDM model, the value of $\alpha$ is indeed very close to $-1$ over a very wide mass range, as shown in Fig.\ref{fig-alpha}. The assumption that $P(m_o\vert m_t)$ can be written as a function of $m_t-m_o$ is also quite commonly adopted in practice \citep{2005PhRvD..72d3006L,2006ApJ...648..956S,2011PhRvD..83b3015M,2013ApJ...767...92R,2017A&A...605A..70D}. For example, in \cite{2008ApJ...676..248Y}, it is found that the luminosity distribution for the galaxy sample has a lognormal form with a scatter nearly independent of the halo mass\footnote{From Fig.\ref{fig-alpha}, we notice that the slope $\alpha(M)$ decreases to -1.3 for $M_t\sim10^{14.3}M_{\odot}/h$. So when a large $\sigma_{lgM}$ shows up, bias would be brought to the high mass end of the RHMF.  To ensure 5 percent accuracy of this method when reconstructing the RHMF, it is better to only use halos whose masses are less than $10^{14}M_\odot/h$ when $\sigma_{lgM}$ achieve 0.35.}.

\begin{figure}
    \centering
    \subfigure{
    \includegraphics[width=0.5\textwidth, clip]{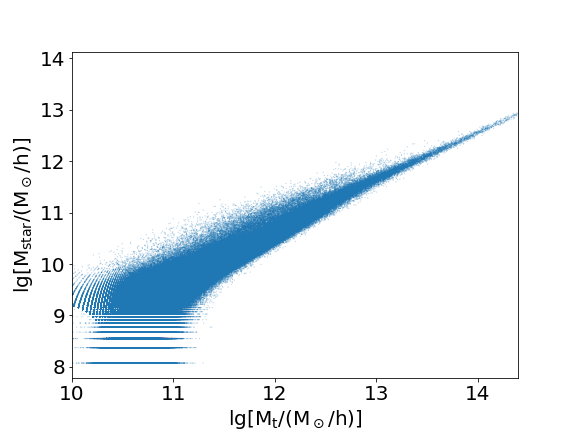}}
    \caption{ The relation between the stellar mass $M_{star}$ and the halo mass $M_t$ for halos from the GadgetMusic simulation of The Three Hundred project\citep{2018MNRAS.480.2898C}.}
    \label{fig-scatter-star}
\end{figure}

\begin{figure}
    \centering
     \subfigure{
    \includegraphics[width=0.49\textwidth, clip]{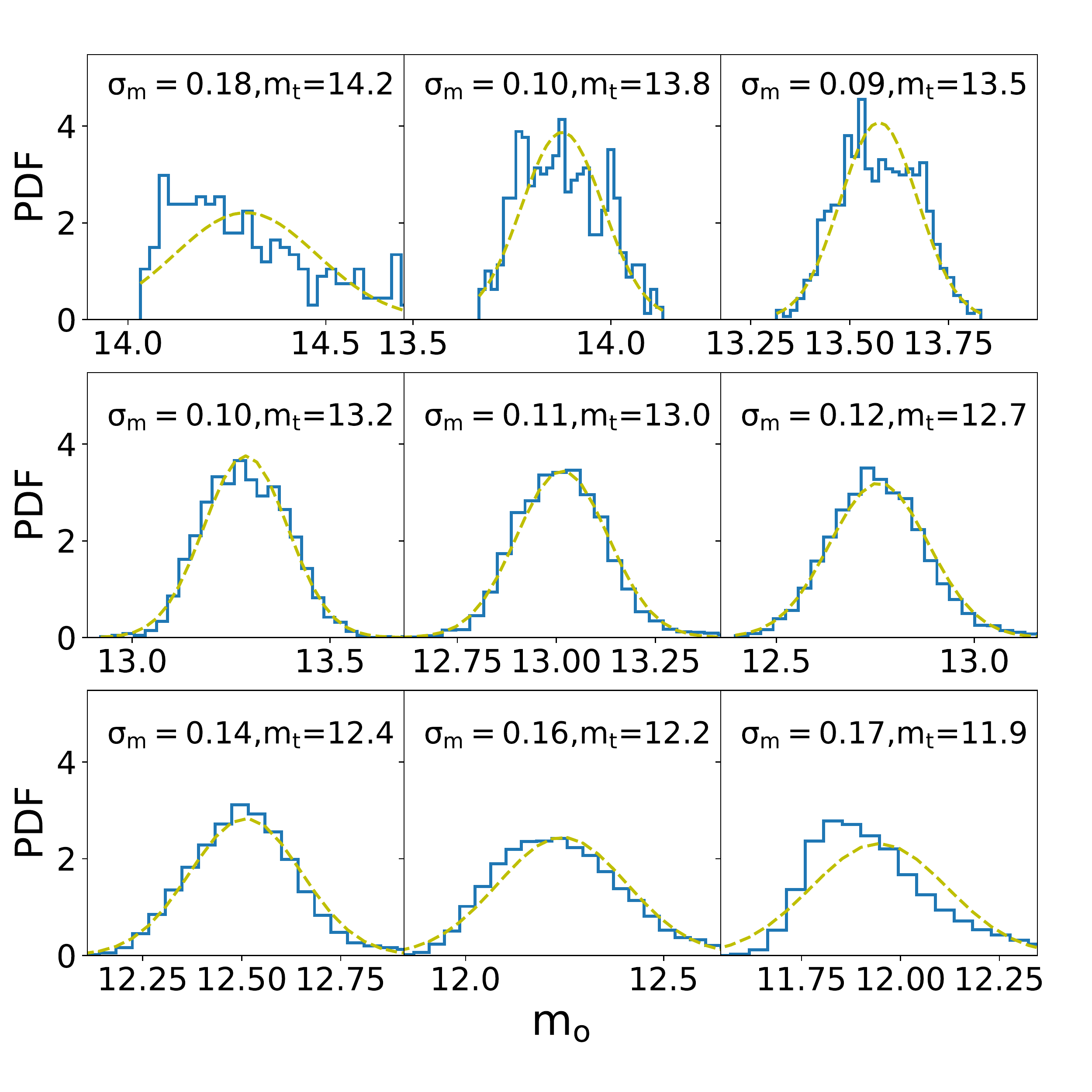}}
    \caption{ \label{fig-pdf-star} 
    The probability distribution (PDF) of $m_o$ (lg$M_o$) for a given $m_t$ (lg$M_t$). The blue histogram shows the PDF of $m_o$ converted from $m_{star}$ shown in Fig\ref{fig-alpha}. The yellow line is the best-fitting gaussian function for the PDF.  $\sigma_m$ is the dispersion of the PDF.
    }
\end{figure}

\begin{figure}
    \centering
    \subfigure{
    \includegraphics[width=0.5\textwidth, clip]{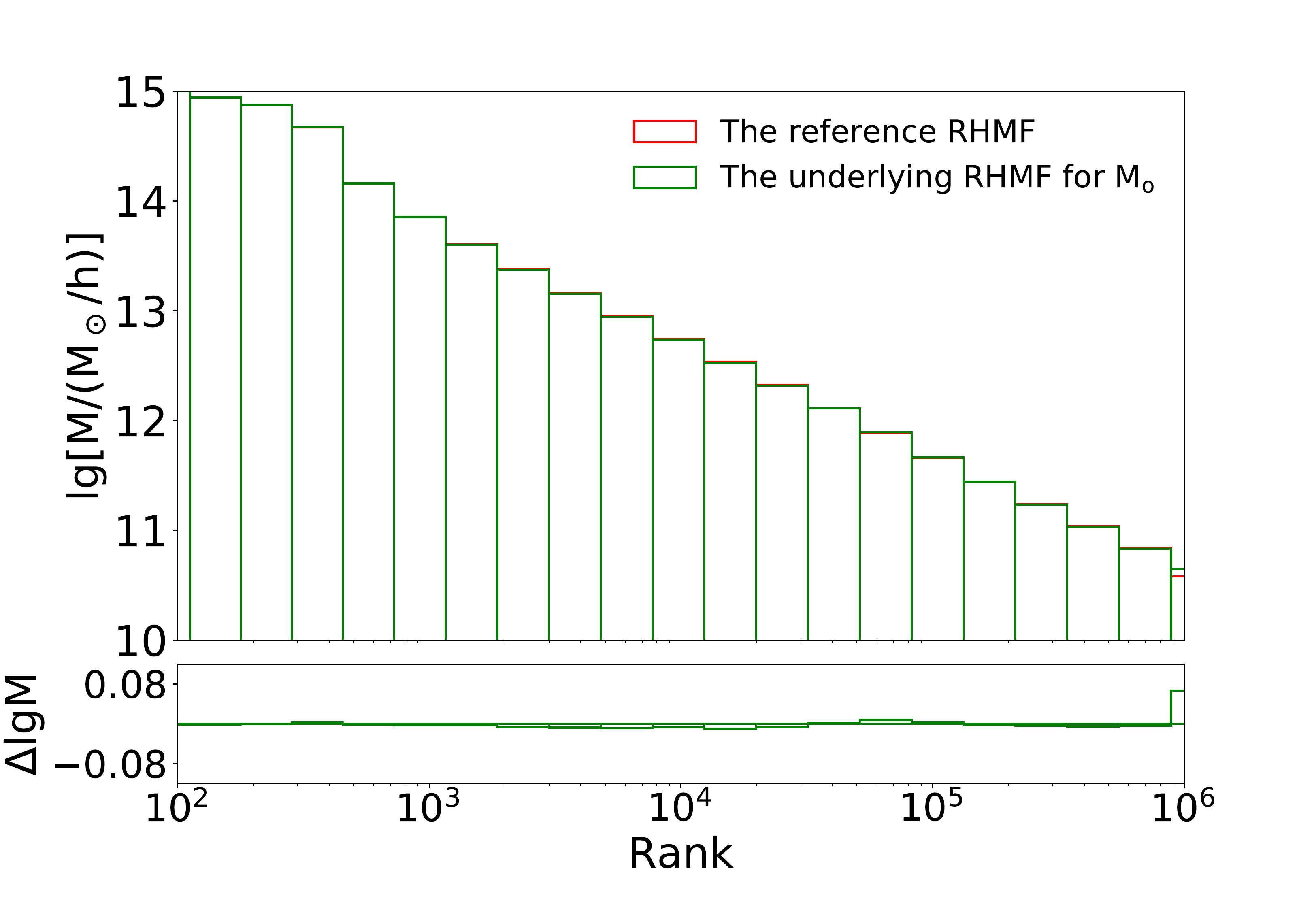}}
    \caption{ The resulting RHMF for halos from the GadgetMusic simulation. The red histogram is the reference RHMF.  The green histogram is the underlying RHMF for $M_o$.}
    \label{fig-hmf-star}
\end{figure}

However, there are also literatures assuming a varying $\sigma_m$ \citep{2015MNRAS.454.2305S,2018ApJ...854..120M}.
For instance, different from \cite{2008ApJ...676..248Y}, \cite{Zheng_2007} find that the scatter of the lognormal luminosity distribution increase from 0.13 for halos with $M_t \sim 10^{13.5}M_\odot/h$  to 0.3 for low-mass halos with $M_t\sim10^{11.5} M_{\odot}/h$.
Therefore it is still important to check the validity of this
assumption for realistic observables. As an example, in
Fig.\ref{fig-scatter-star}, we plot $m_{star}$ ($\lg M_{star}$)
vs. $m_t$ ($\lg M_{t}$) for halos from the GadgetMusic simulation,
which is one of the zoom-in hydro-simulations from the Three Hundred project \citep{2018MNRAS.480.2898C}. By binning the halos according to their true masses, we derive the average relation between $m_t$ and $m_{star}$ in logarithmic space, and use the relation to convert each $m_{star}$ into a halo mass $m_o$. In Fig.\ref{fig-pdf-star}, we show $P(m_o\vert m_t)$  for different halo masses $m_t$, and fit each of them with a Gaussian function. The value of $\sigma_m$ is relatively stable when $m_t\gsim 13$, but increases gradually for smaller halo masses. Nevertheless, even in this case, we find that the underlying RHMF by sorting $m_o$ is still quite close to the reference RHMF, as shown in Fig.\ref{fig-hmf-star} with the green histogram (the definition of the histograms are the same as those in Fig.\ref{fig-hmf-rand}). It implies that the accuracy of RHMF by sorting observables is not very sensitive to the variation of $\sigma_m$!

\begin{figure}
    \centering
    \subfigure{
    \includegraphics[width=0.5\textwidth, clip]{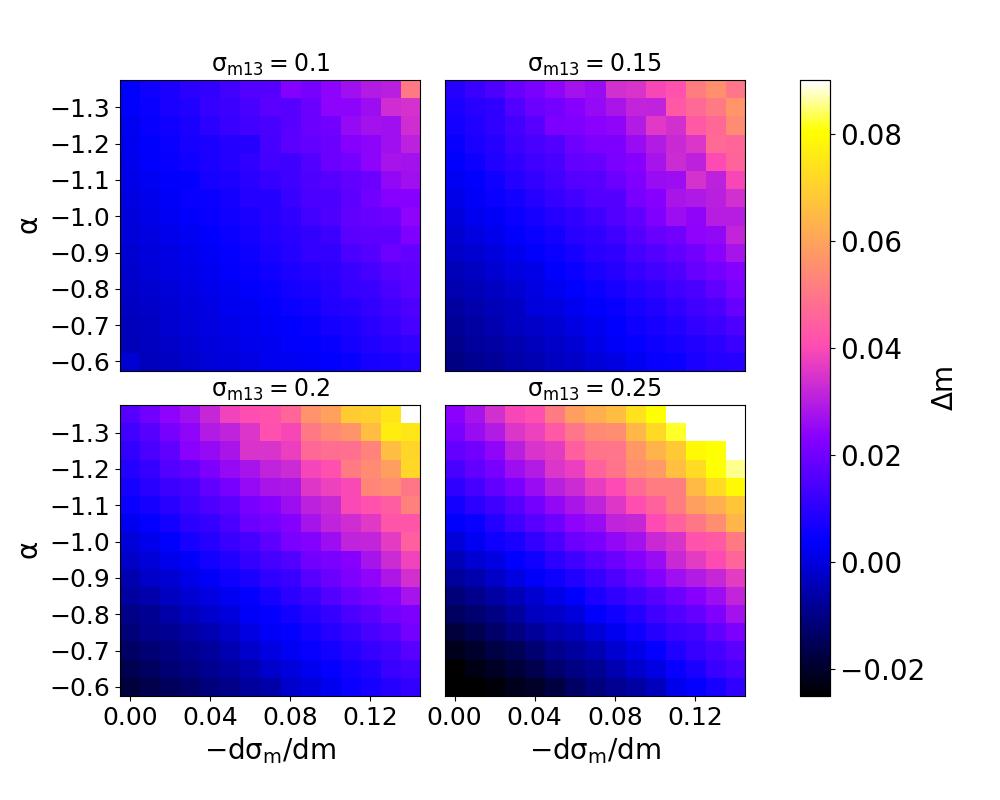}}
    \caption{\label{fig-slope} 
    The deviation of the underlying RHMF for $M_{o}$ from the reference RHMF. It is shown for different $\sigma_{m13}$, $d\sigma_m/dm$ and the power index $\alpha$. For each panel, the x-axis is the $-d\sigma_m/dm$, and the y-axis is the power index $\alpha$ of the halo mass function.  The case shown in Fig.\ref{fig-hmf-star} corresponds to $\alpha\sim -0.9$, $-d\sigma_m/dm\sim -0.04$, and $\sigma_{m13}=0.11$.}
\end{figure}

More generally, we carry out a series of tests to check the bias of RHMF for $d\sigma_{m}/dm\ne 0$ and $\alpha\ne-1$. For simplicity, we define the bias of the RHMF as the median of the relative mass biases in different bins. Since the bias also depends on the overall magnitude of $\sigma_m$, we use $\sigma_{m13}$ to denote the scatter at $M_t=10^{13}M_\odot/h$.    
In Fig.\ref{fig-slope}, we show how the bias of RHMF changes with $d\sigma_m/dm$ and $\alpha$ for several different choices of $\sigma_{m13}$. The results indicate that the bias takes very small values ($\lsim 0.02$) over a wide range of parameter space, demonstrating the robustness of the sorting method in constructing RHMF. In our earlier example from the GadgetMusic simulation, we have $\alpha\sim -0.9$, $d\sigma_m/dm\sim -0.04$, and $\sigma_{m13}=0.11$, which indeed corresponds to a very small bias according to Fig.\ref{fig-slope}.



\section{Mass Measurement with Weak Lensing}
\label{simulation}

We have shown in the previous section that the RHMF can be accurately recovered by binning the observables.
In this section we will show that the $\overline{M}_t$ of each bin can be measured with weak lensing. 
Firstly, we will introduce the mass measurement with the lensing signals generated by the
ideal Navarro-Frenk-White halos (hereafter NFW) \citep{1996ApJ...462..563N}. 
And then, we will show the mass measurement with halos in the N-body simulations.


\subsection{Spherical NFW Excess Surface Density}
\label{section-NFW}

Assuming the halos have spherical NFW profiles on average, the common way to get the halo masses is to fit the lensing signals 
around the halos with the NFW excess surface density (hereafter ESD) \citep{1996A&A...313..697B,2000ApJ...534...34W}.

For the spherical NFW halos, their 3D mass density are given by: 
\begin{equation}
\rho(r)=\frac{\rho_c}{r/r_s(1+r/r_s)^2},
\end{equation}
where $\rho_c$ and $r_s$ are two parameters that describe the density profile. 
We parametrize the halo with the viral mass $M_{vir}=(4\pi/3)r_{vir}^3\rho_{vir}$
and the concentration $c_{vir}=r_{vir}/r_s$, where $\rho_{vir}=\Delta_{vir}\rho_{crit}$.
In the $\Lambda CDM$ model, the $\Delta_{vir}$ can be approximately calculated as \citep{1998ApJ...495...80B}
\begin{eqnarray}
\label{eq-mvir}
\Delta_{vir}=18\pi^2+82[\Omega_m(a)-1]-39[\Omega_m(a)-1]^2.
\end{eqnarray}
The viral radius $r_{vir}$ is defined as the radius within which the mean mass density of the halo
reach $\rho_{vir}$. $\rho_c$ is given by:
\begin{eqnarray}
\rho_c=\frac{\rho_{vir}}{3}\frac{c^3}{ln(1+c)-c/(1+c)}
\end{eqnarray}

We define z as the distance along the line of sight, and R as the projected comoving radius from halo center, with $r=\sqrt{R^2+z^2}$. Introducing a dimensionless quantity $x=R/r_s$, the projected surface density of the halo:
\begin{equation}
\Sigma(x)=\int^{+\infty}_{-\infty}\rho(r_s,x,z)dz,
\end{equation}
and the mean surface density inside the radius R is
\begin{equation}
\label{Sigmax}
\Sigma(<x)=\frac{1}{\pi x^2}\int^x_02\pi x\Sigma(x)dx.
\end{equation}
The ESD for the NFW halo is given by
\begin{eqnarray}
\label{eq-excess}
\Delta\Sigma(R)=\Sigma(<x)-\Sigma(x)=2\rho_cr_sf^{NFW}(x), 
\end{eqnarray}
where the function $f^{NFW}$\citep{2015PASJ...67..103N} is given by 
\begin{eqnarray}
\label{fnew}
&&f^{NFW}(x)= \\\nonumber
&&\left\{\begin{array}{ll}\frac{2}{x^2}ln\frac{x}{2}+\frac{1}{1-x^2}\left(1+\frac{2-3x^2}{x^2\sqrt{1-x^2}}cosh^{-1}\frac{1}{x}\right),(x<1) \\ 
\frac{5}{3}-2ln2,\;\;\;\;\;\;\;\;\;\;\;\;\;\;\;\;\;\;\;\;\;\;\;\;\;\;\;\;\;\;\;\;\;\;\;\;\;\;\;\;\;\;\;\;\;\;(x=1) \\
\frac{2}{x}ln\frac{x}{2}-\frac{1}{x^2-1}\left(1+\frac{2-3x^2}{x^2\sqrt{x^2-1}}cos^{-1}\frac{1}{x}\right),\;\;\;(x>1) \end{array}\right.
\end{eqnarray}

\subsection{Measuring the RHMF with the ideal NFW halos}
\label{section-NFW-lognormal}

\begin{figure}
    \centering
    \subfigure{
    \includegraphics[width=0.5\textwidth, clip]{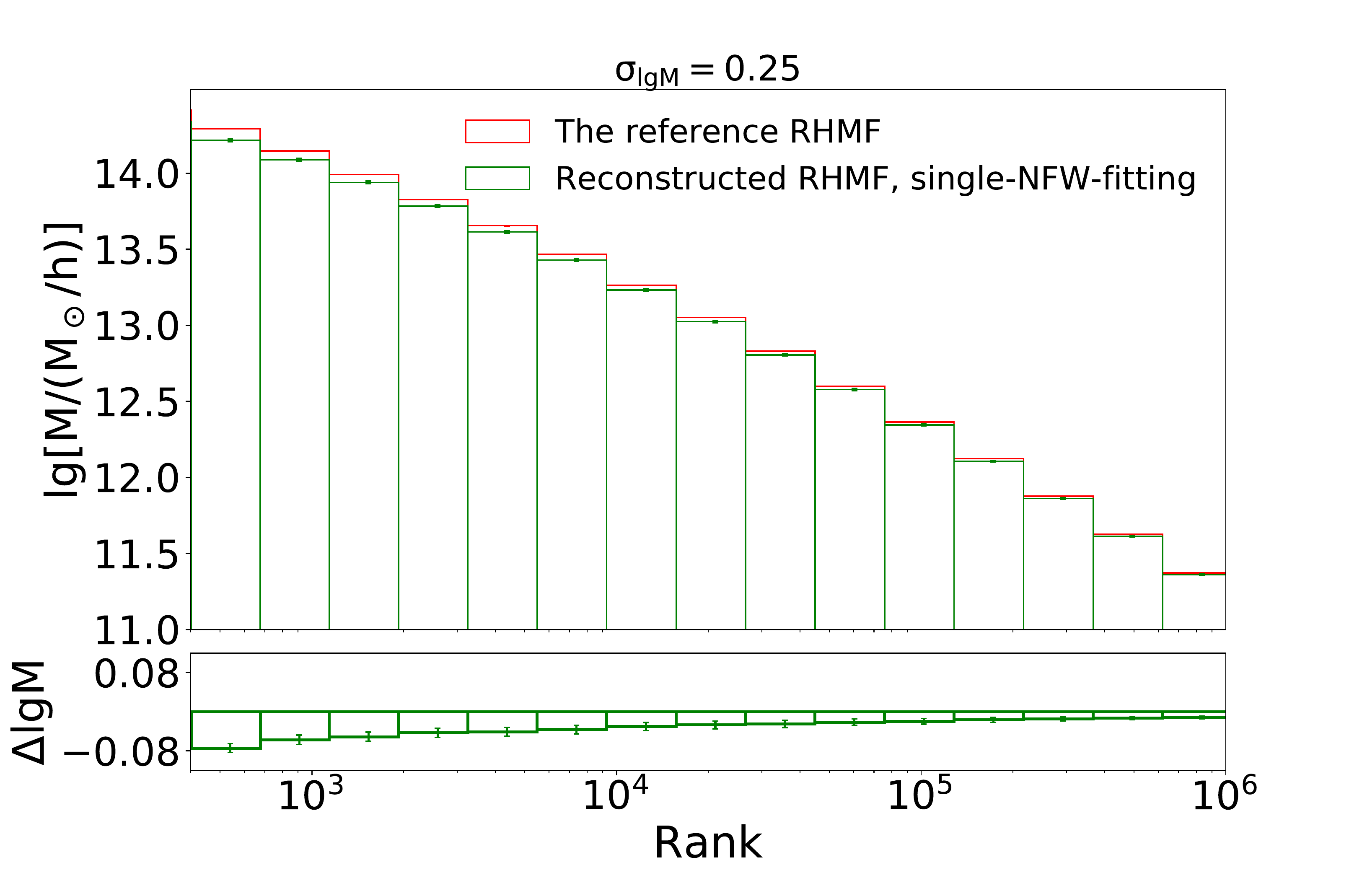}}
    \caption{\label{fig-single-hmf-lens} 
    The green histogram is the reconstructed RHMF for halos sorted by $M_{o}$, with the mass of each bin recovered by fitting the stacked ESD using that of a single NFW profile. The red histogram is the reference RHMF for comparison.}
\end{figure}

\begin{figure*}
    \centering
    \subfigure{
    \includegraphics[width=1\linewidth, clip]{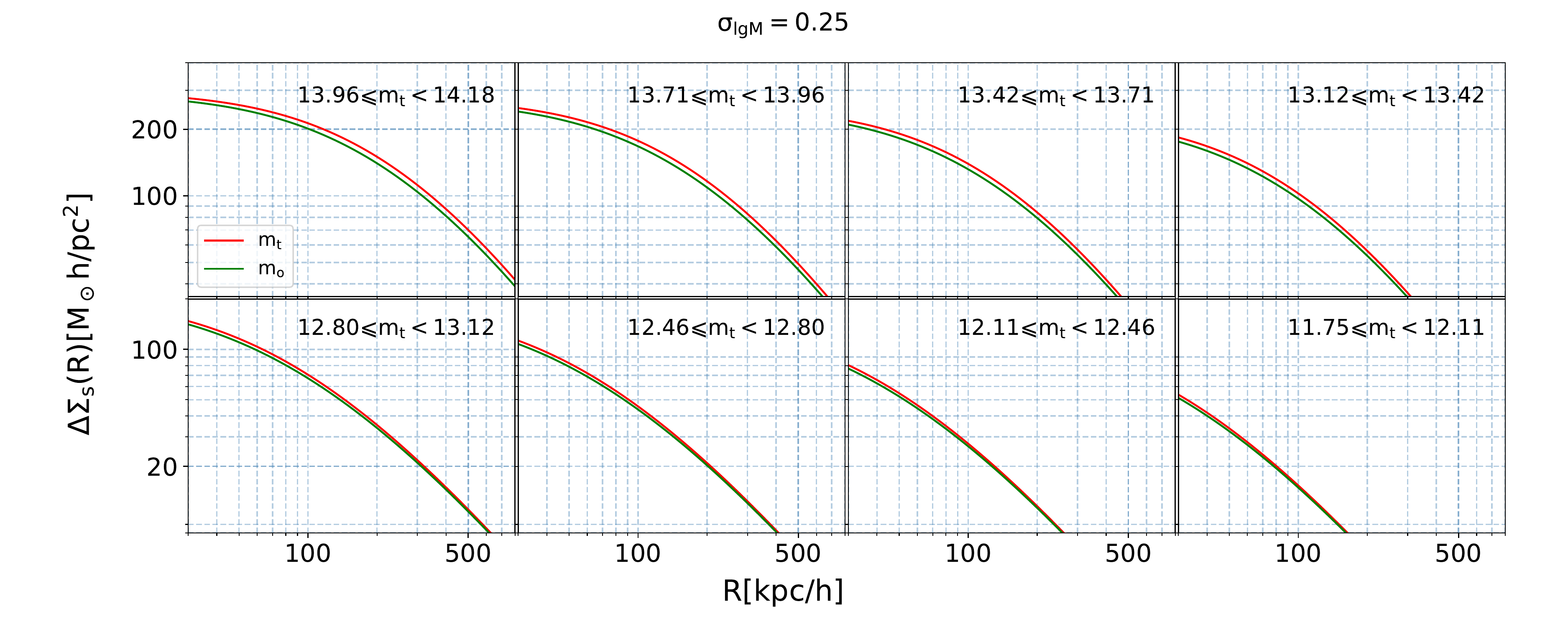}}
    \caption{\label{fig-150-profile} 
    Comparison of the stacked ESD profiles for bins that are sorted by $M_t$ (in red) and $M_o$ (in green ) respectively. }
\end{figure*}

To test the idea of \S\ref{method}, we generate a set of halos according to the S-T mass function, and assign to each halo a spherical NFW density profile with concentration given by the m-c relation from \cite{2009ApJ...707..354Z}. A mass $M_{o}$ (assuming it is converted from an observable) is generated by adding a lognormal error to each $M_{t}$, with $\sigma_{{\mathrm lgM}}=0.25$. We rank the halos by their $M_{t}$ and $M_{o}$ respectively, and form the bins for measuring the RHMF. It is the purpose of this section to find out how well the underlying RHMF from the bins formed by $M_{o}$ can be recovered by weak lensing, and achieve a good comparison with the results from bins formed by $M_{t}$. 

If we fit the stacked profile of halos in each bin with the ESD from a single NFW profile, we find that the results from bins generated with $M_{t}$ and $M_{o}$ can be somewhat different, as shown in Fig.\ref{fig-single-hmf-lens}. This is perhaps not surprising, as the ESD is not a linear function of the halo mass. The conclusion of \S\ref{method} therefore cannot be extended to the case of the stacked ESD. This fact is further shown in Fig.\ref{fig-150-profile}, in which the red and green profiles are the stacked profiles from $M_{t}$ and $M_{o}$ respectively. Their differences are indeed quite obvious.  In other words, although the halos sorted by $M_t$ and $M_o$ have the same average mass in each bin, they do not share the same average ESD profile.

\begin{figure}
    \centering
    \subfigure{
    \includegraphics[width=0.5\textwidth, clip]{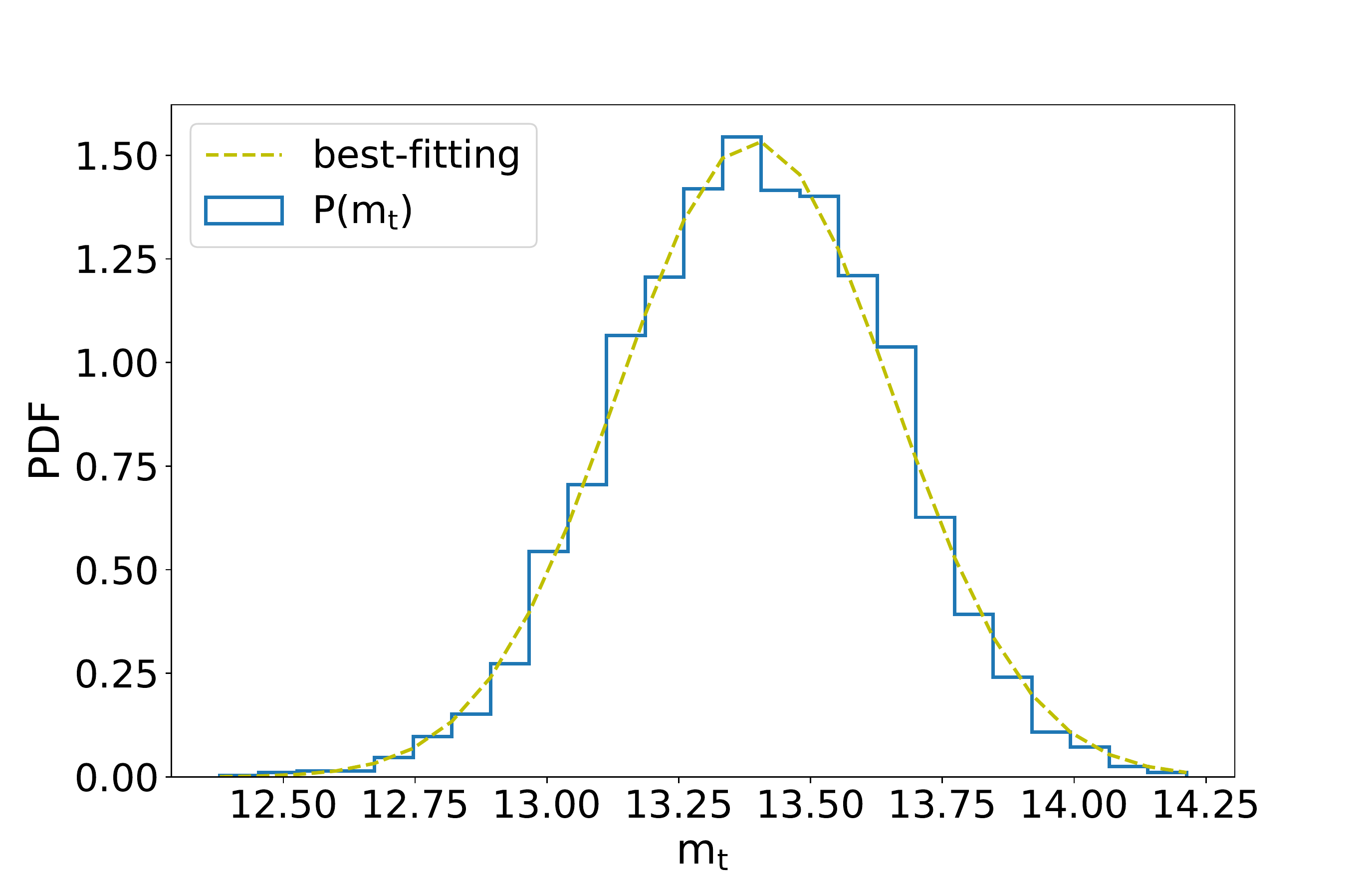}}
    \caption{\label{fig-idealNFW-pdf} 
    The distribution of $m_t$ ($lgM_t$) for a bin formed by $13.44<m_o<13.64$. The blue histogram shows the actual mass distribution $P(m_t)$. It is well approximated by a lognormal distribution shown as the yellow curve.}
\end{figure}

\begin{figure}
    \centering
    \subfigure{
    \includegraphics[width=0.5\textwidth, clip]{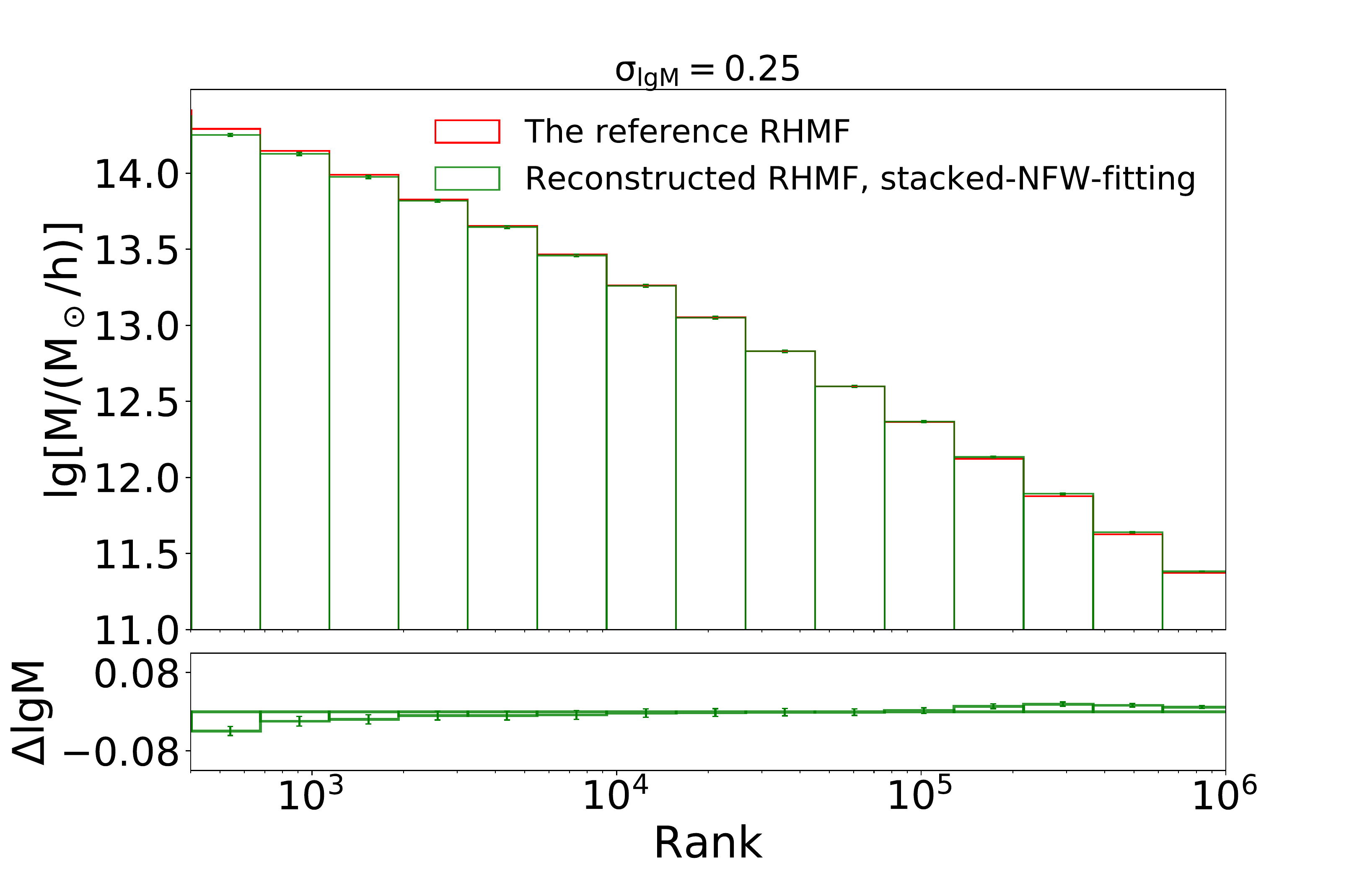}}
    \caption{\label{fig-gauss-hmf-lens} 
    The measured RHMF with the stacked-NFW-fitting for the ideal NFW halos. 
    The red histogram is the reference RHMF, and the green histogram shows the reconstructed RHMF for $M_o$.}
\end{figure}

To avoid such a bias, we introduce the stacked-NFW-fitting to take into account the distribution of the true halo masses within each bin defined by $M_{o}$. The idea is to fit the stacked ESD from observation with the stacked model ESDs for halo masses generated by a certain mass distribution. The best fitting case determines the parameters of the mass distribution, which in turn yields the arithmetic mean of the halo mass for the bin. The form of the mass distribution can in principle be worked out given the statistical relation between $M_{o}$ and $M_{t}$. However, to a good approximation, we find it convenient to assume that the true halo mass within each bin simply follows a lognormal distribution. The central value and the dispersion of the distribution are the only two parameters for the fitting. As an example, in Fig.\ref{fig-idealNFW-pdf}, we show the mass distribution of $m_t$ for a bin defined by $13.44<m_o<13.64$ (in blue). It is well approximated by a lognormal distribution shown as the yellow curve in the figure.
The resulting RHMF is shown in Fig.\ref{fig-gauss-hmf-lens}. As presented in the upper panel of the figure, the measured underlying RHMF with $M_{o}$ is consistent with the reference RHMF. The residuals are found to be small as shown in the lower panel.

In summary, the underlying mean masses $\overline{M}_{t}$ for bins sorted by $M_{o}$ can be better measured with the stacked-NFW-fitting method. Fitting with the ESD of a single NFW can cause systematic biases due to the nonlinear relation between the halo mass and the ESD. In the next section, we turn to halos in N-body simulations to further check the accuracy in the recovery of RHMF, and identify problems.


\subsection{Measuring the RHMF with N-body simulations}
\label{section-simulation}

Our N-body simulation \citep{2007ApJ...657..664J} assumes the $\Lambda$CDM cosmology, with $\Omega_c=0.268$, $\Omega_b=0.045$, $\Omega_\Lambda=0.732$, $\sigma_8=0.85$, $h=0.71$, and $n_s=1$. The simulation uses $1024^3$ particles in a box of size $300 {\mathrm Mpc/h}$, with the particle mass equal to $0.187\times 10^{10}\mathrm{M_{\odot}/h}$. Halos are identified using the standard FoF algorithm with linking length equal to 0.2. The subhalos are identified with the HBT algorithm\citep{2012MNRAS.427.1651H}. The most bound particle of the central subhalo are chosen to be the halo's central position, in order to avoid the center offset effect. Considering the simulation resolution, only the FoF halos with particle number $NP>100$ are used. We use the mass $M_{fof}$ from the Friends-of-Friends method to denote the halo mass $M_{t}$. The mass $M_{o}$ from observable is generated by adding a lognormal error to $M_{t}$ with $\sigma_{\mathrm lgM}=0.25$.

\begin{figure}
    \centering
    \subfigure{
    \includegraphics[width=0.5\textwidth, clip]{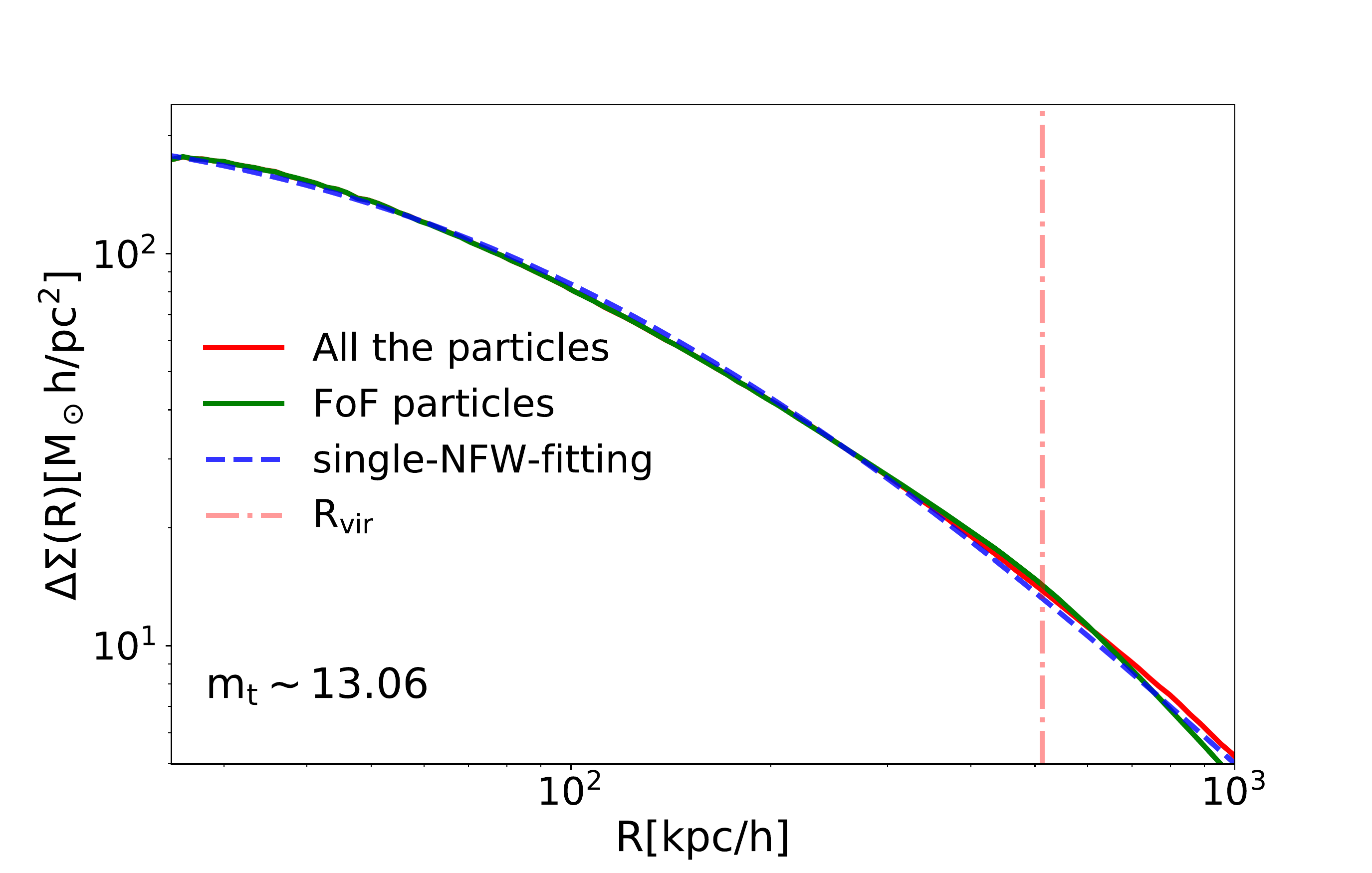}}
    \caption{\label{fig-simu-prof} 
    The stacked ESD for halos with masses $m_t$ ($lg[M_t/(M_\odot/h)]$) in a narrow mass bin. The masses for these halos are 
    around 13.06, and the mass bin size is about 0.15 dex. The red solid line shows the ESD achieved with all the
    particles within the slice of projection. The green solid line shows the ESD achieved with the particles belonging 
    to these FoF halos. These two solid lines begin to separates only when the projected radius is larger than $R_{vir}$. 
    The blue dashed line is the best-fit ESD of the NFW profile.}
\end{figure}

For our purpose, it is useful to first study the profile of the stacked ESD for the dark matter halos \citep{2006MNRAS.373.1159Y}. This is done by projecting the positions of the dark matter particles of halos onto a plane perpendicular to the line of sight. Fig\ref{fig-simu-prof} shows the stacked ESD for halos with masses $M_{t}$ in a narrow mass bin ($\Delta m \sim$ 0.15 dex) around $10^{13.06}M_\odot/h$. The red curve is achieved with all the particles within the slice of projection \footnote{ The ESD converges well for the width of the slice being larger than $10 Mpc/h$. }. The green curve is calculated from the particles belonging to the FoF groups of halos. Note that only when the projected distance to the halo center exceeds the viral radius ($R_{vir}$), the two curves begin to deviate from each other, indicating that within $R_{vir}$, it is safe to fit the ESD without considering the 2-halo term. 

The blue dashed curve in Fig\ref{fig-simu-prof} shows the best-fit ESD from a single NFW profile. We are aware that there are some scale-dependent residuals in the fitting. With the stacked-NFW-fitting method introduced in \S\ref{section-NFW-lognormal}, we find that if only using the average ESD curves between [23 kpc/h, 0.5 $R_{vir}$], the reference RHMF could be well recovered.
However, we find that this mass measurement  is influenced by the fitting range, for which the detailed discussion is given in appendix\ref{appendix}.



The quality of stacked-NFW-fitting is limited by the accuracy of the NFW profile. One may consider other empirical models for describing the halo density profiles, such as gNFW\citep{1996MNRAS.278..488Z,Jing_2000}, Einasto\citep{1965TrAlm...5...87E}, BMO\citep{2009JCAP...01..015B}, Prugniel-Simien \citep{1997A&A...321..111P,Merritt_2006}, etc.. This is however not necessary. Instead, we choose to create the model ESD directly from the N-body simulation. The method ``stacked-NFW-fitting'' should therefore be more generally called ``stacked-model-fitting''. We use halos from another independent N-body simulation to produce the ESD model as a function of the halo mass. The new ESD models can in principle be made more isotropic by averaging them over several different line-of-sight directions. However, we find that even this is not really necessary in the stacked-fitting, as stacking itself removes a significant amount of anisotropy in the final ESD model. 

\begin{figure}
    \centering
    \subfigure{
    \includegraphics[width=0.5\textwidth, clip]{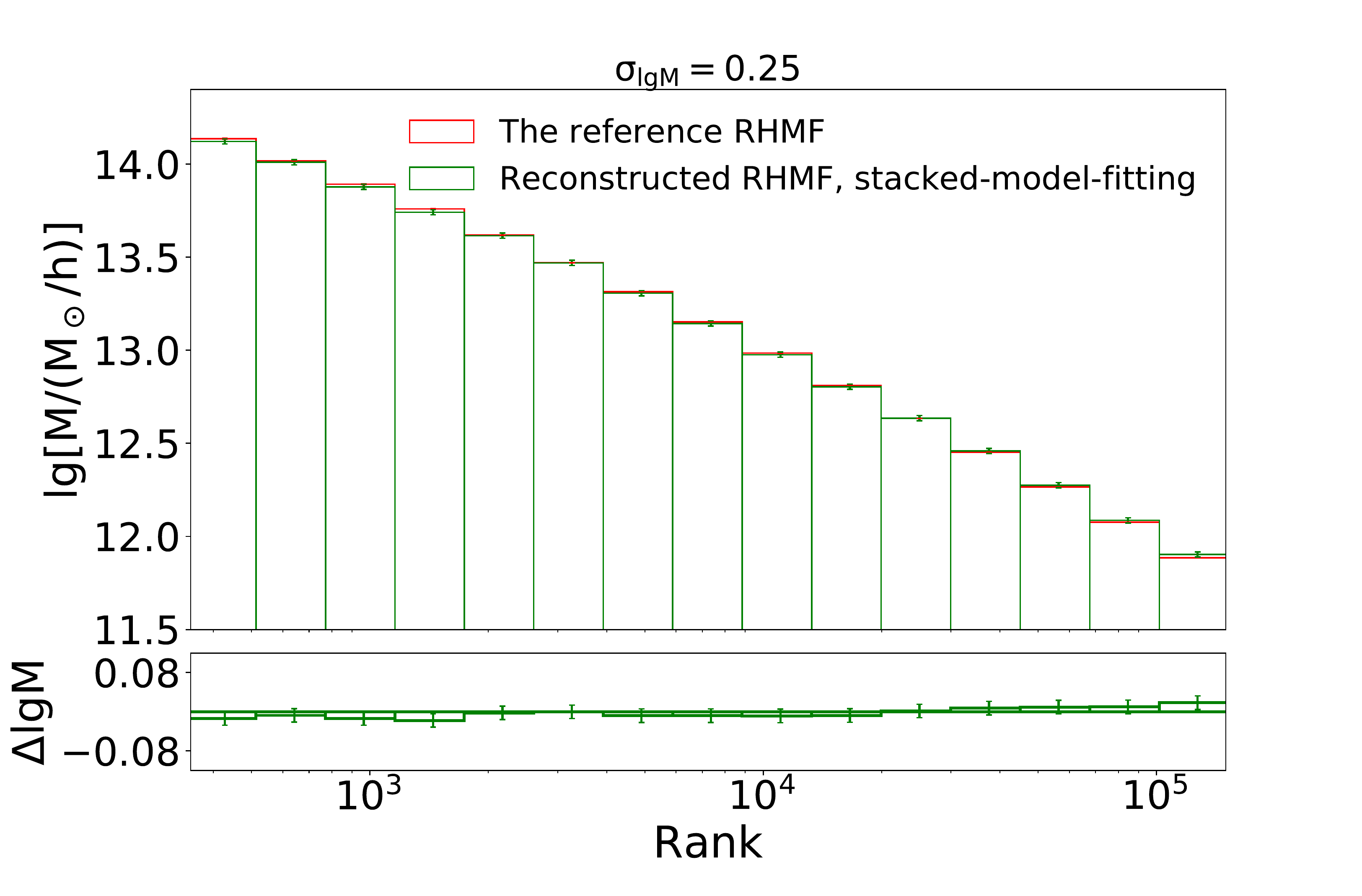}}
    \caption{\label{fig-template} 
    The green histogram shows the RHMF measured with the stacked ESD model created from the N-body simulation for bins formed by $M_o$. The red histogram is the reference RHMF.}
\end{figure}

Similar to what is done in stacked-NFW-fitting, we assume that the true halo mass distribution in each bin ordered by $M_{o}$ follows a lognormal distribution. The mean and dispersion of the distribution are achieved by fitting the stacked ESD with those stacked by the ESD models. The RHMF measured in this way is shown as the green histogram in Fig.\ref{fig-template}. It agrees very well with the reference RHMF (shown in red). 

\section{Conclusion and Discussions}
\label{discussion}

Accurate measurement of the halo mass function is an important subject in precision cosmology. A main concern in this field is about building up the relation between the obervable and the underlying halo mass. The dispersion between the observable and the halo mass is also required to be known well to overcome the Eddington bias in traditional methods. In this paper, we point out a new way of measuring the halo mass function that does not require accurate knowledge of the mass-observable relation. In the new method, the observable is used only for the purpose of sorting the halos as if the observable were the true halo mass. It is to our surprise that the resulting mean halo mass of each bin is indeed very close to its counterpart from sorting the true halo mass! In \S\ref{method}, we have shown that this phenomenon relies on two conditions: 1. the halo mass function is close to a power-law form with a power index close to $-2$, \ie, the total halo mass within each unitary logarithmic scale is approximately a constant (This condition is approximately satisfied by the $\Lambda$CDM models over a large mass range); 2. statistically, the observable has a monotonous relation with the halo mass, and the dispersion of the relation changes slowly with the halo mass.

Making use of this appealing feature, we show how to use weak lensing to accurately recover the so-called Ranked Halo Mass Function (RHMF), \ie, the halo mass as a function of the mass order/rank within a certain cosmic volume. This can be done by modeling the true halo mass distribution as a lognormal function in each bin defined by the observable, and fitting their stacked excess surface density profile (ESD) with stacked ESD models from either NFW or N-body simulations directly. In this way, halo mass function from the cosmological simulation can be directly compared with that from observations without careful modelling of the mass-observable relation!

\begin{figure*}
    \centering
    \subfigure{
    \includegraphics[width=1\textwidth, clip]{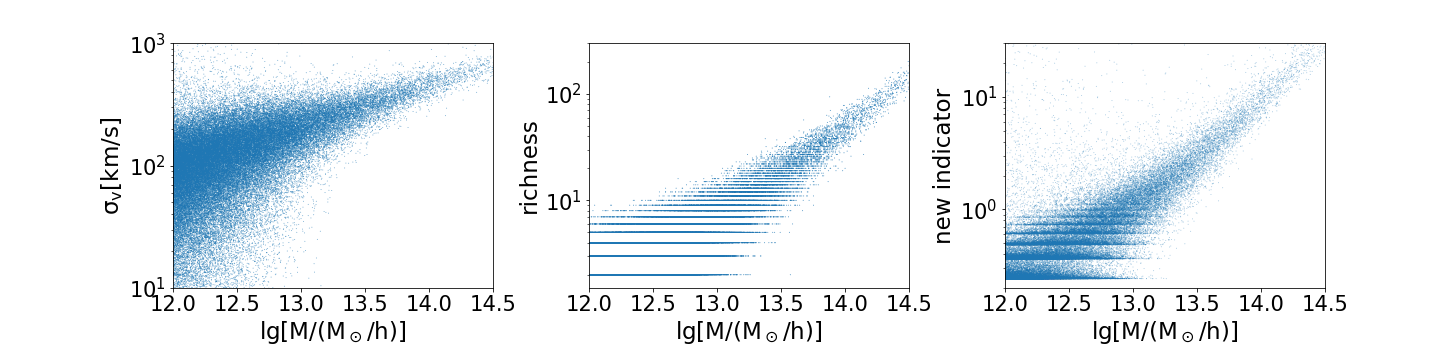}}
    \caption{\label{fig-sigma-rich}The halo mass vs. observables for the N-body simulation.
     The left panel shows halo mass vs. line-of-sight velocity dispersion ($\sigma_v$). The middle panel shows the halo mass vs. richness (r).  The right panel shows the halo mass vs. the form of $r/r_0+2\sigma^4_v/\sigma^4_{v0}$, with $r_0$ and $\sigma_{v0}$ being constants.
     }
\end{figure*}

\begin{figure}
    \centering
    \subfigure{
    \includegraphics[width=0.5\textwidth, clip]{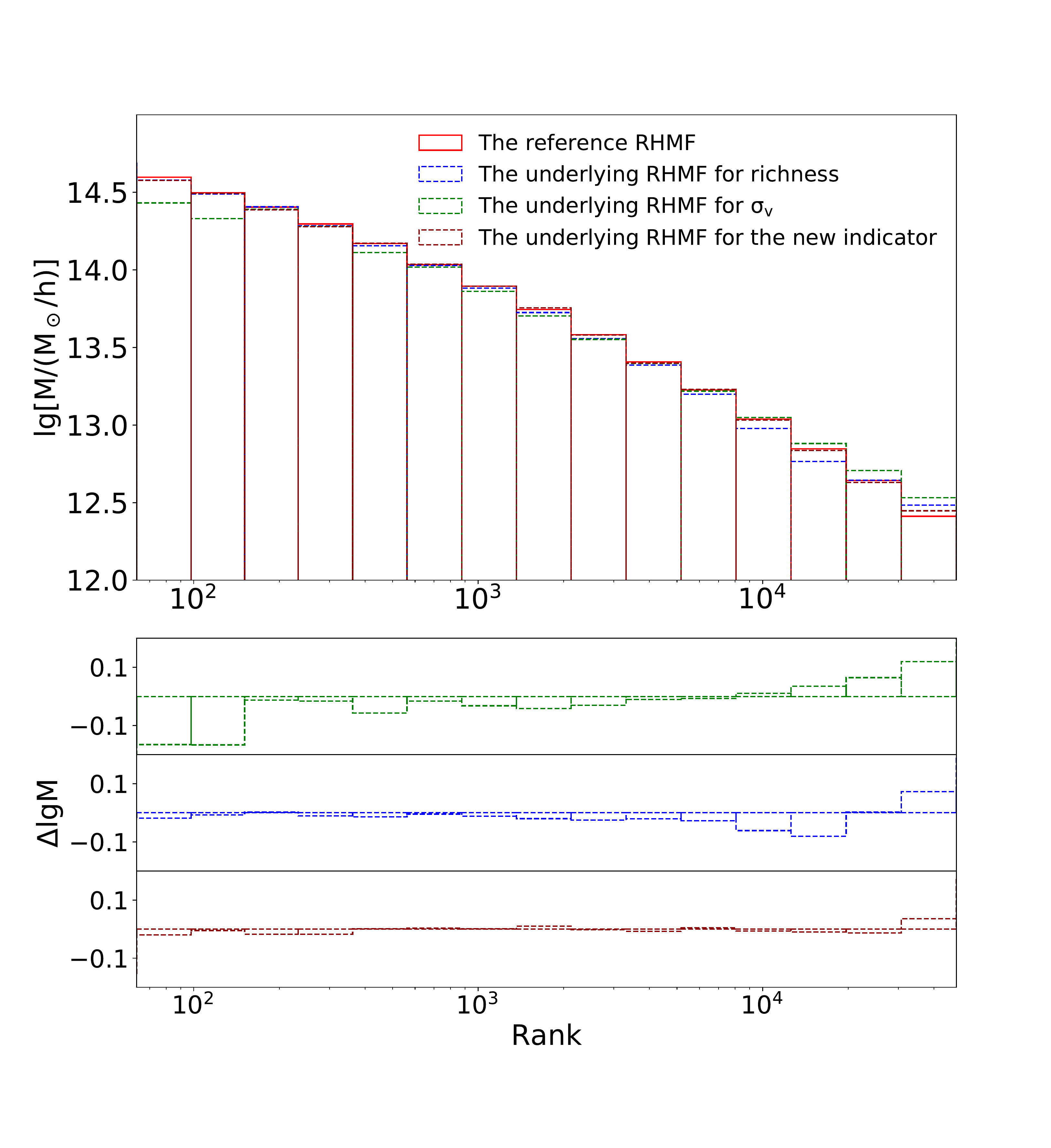}}
    \caption{\label{fig-Mtrue-combine} 
    The resulting RHMF for the observable. The red solid histogram is the reference RHMF. The blue and green dashed histogram show the underlying RHMFs for richness and $\sigma_v$ respectively. The marroon dashed histogram show the underlying RHMF for the mass indicator defined as a function of $\sigma_v$ and richness.}
\end{figure}

We point out that although the study in this work uses the Friends-of-Friends mass to define the halo mass, other mass definitions such as $M_{vir}$, $M_{200}$, and $M_{500}$ are equally good as long as the same mass definition is used for both constructing the reference RHMF and creating the ESD models, because these definitions have very tight relations with each other. Note that changing the mass definition leads to a change in the halo mass function and the reference RHMF, meanwhile the constructed RHMF from the observables changes correspondingly. The mass definition therefore does not affect the comparison of the RHMF's from observation and simulation. We consider this feature a significant advantage of our method. In fact, this is in line with the main idea of this paper. If we regard the different mass definitions as different observables, they shall share the same underlying RHMF as long as the dispersions of their mutual relations are moderate . It can be viewed as an option to achieve precise cosmological constraints without worrying too much about the robustness of the halo mass definition.

For accurate reconstruction of the RHMF, it is better to have an observable (used as the mass estimator) that has a tight relation (less scatter) with the halo mass. For this purpose, one can consider building an observable as a function of the existing/familiar ones. For instance, the subhalo velocity dispersion $\sigma_v$ along the line of sight is found to have large scatters with the halo mass $M_t$, as shown in the left panel of Fig.\ref{fig-sigma-rich}, in which many small halos are found to have great $\sigma_v$. The situation for richness\footnote{ We define the halo richness as the number of its subhalos with masses lager than $10^{10.4}M_\odot/h$.} $r$ is somewhat similar, as shown in the middle panel of Fig.\ref{fig-sigma-rich}. As an improvement, we consider the combination of these two quantities in the form of $r/r_0+2\sigma^4_v/\sigma^4_{v0}$, which seems to have more tight relations with $M_t$, as shown in the right panel of Fig.\ref{fig-sigma-rich}.  Here, the $r_0$  is set to be the value of richness at $m_t=13$  using the average relationship between the richness and the halo mass. $\sigma_{v0}$ is similarly defined, but with $m_t=13.5$. Its underlying RHMF becomes more consistent with the reference RHMF with this new mass indicator, as shown in Fig.\ref{fig-Mtrue-combine}.

{In construction the RHMF, there are issues that have not been
discussed in this work. For example, the off-centering effect in the
stacking of the background lensing signals, subhalo contribution of
the satellite galaxies, stellar contribution of the central galaxies
\citep[see e.g.][for more details]{2018ApJ...862....4L}, as well as the baryonic effect on the
halo density profile, the miss-identification of clusters/groups in
galaxy redshift surveys \citep{2007ApJ...671..153Y}. In our future
work, we plan to address these issues when applying this method on the
real data.

\acknowledgments{ACKNOWLEDGMENTS}
\\

We thank Yipeng Jing for providing us the N-body simulations. We also
thank Weiguang Cui for providing us the halo catalogue of The Three
Hundred simulation. This work is supported by the National Key Basic
Research and Development Program of China (No.2018YFA0404504), the
National Key Basic Research Program of China (2015CB857001,
2015CB857002), the NSFC grants (11673016, 11433001, 11621303). JJZ is supported by China
Postdoctoral Science Foundation 2018M632097.
\\
\bibliography{RHMF}

\begin{appendices}
\section{The mass measurement for halos from the N-body simulation with the NFW model}  
\label{appendix}

In \S\ref{section-NFW-lognormal}, for the case of ideal NFW halos, we show that the underlying mean masses for bins sorted by $M_o$ can be better measured with the stacked-NFW-fitting method.  In this section, for halos from the N-body simulation, we perform the mass measurement  in a similar way. The only difference is that we introduce an additional parameter  $c_{ratio}$ in the fitting to account for a mass-independent multiplicative deviation of the concentration from the m-c relation of \cite{2009ApJ...707..354Z}. \\


\begin{figure}
    \centering
    \subfigure{
    \includegraphics[width=0.5\textwidth, clip]{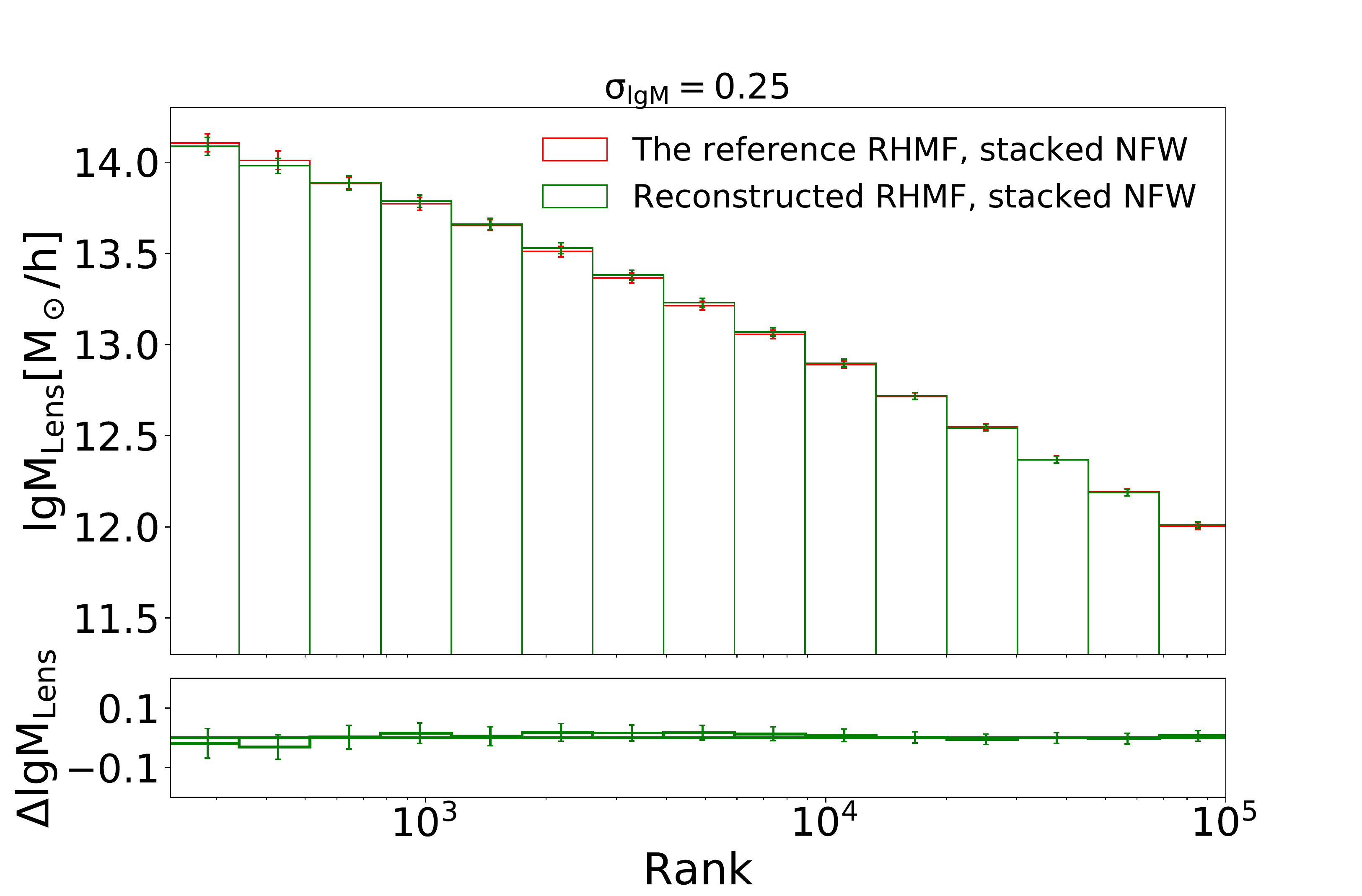}}
    \caption{\label{fig-gauss-hmf-NFW-range1-simu} 
    The measured RHMF with the stacked-NFW-fitting, using the stacked ESD curve between  [23kpc/h, 0.5 $R_{vir}$]. The green histogram shows the reconstructed RHMF for $M_o$. The red histogram is the reference RHMF.}
\end{figure}
\begin{figure}
    \centering
    \subfigure{
    \includegraphics[width=0.5\textwidth, clip]{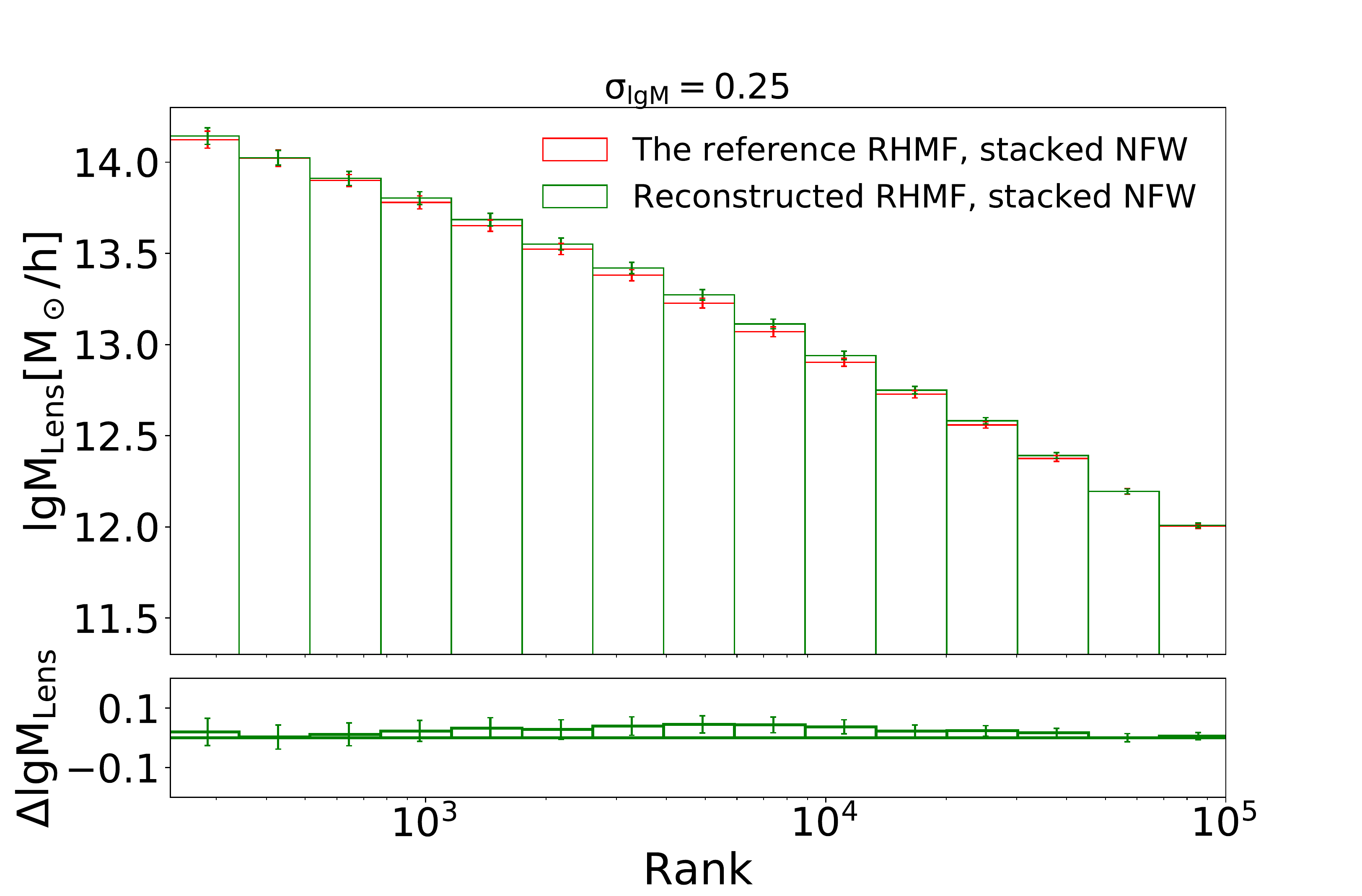}}
    \caption{\label{fig-gauss-hmf-NFW-range2-simu} 
    The measured RHMF with the stacked-NFW-fitting, using the stacked ESD curve between  [23kpc/h, 0.7 $R_{vir}$]. The green histogram shows the reconstructed RHMF for $M_o$. The red histogram is the reference RHMF.}
\end{figure}

With the stacked-NFW-fitting, we find that if only using the stacked ESD curves between [23kpc/h, 0.5 $R_{vir}$], the reference RHMF could be well recovered, as shown in Fig.\ref{fig-gauss-hmf-NFW-range1-simu}. When attempting to extend the fitting range to, say, 0.7 $R_{vir}$, we find that the noticeable inconsistencies show up between the measured RHMF and the reference RHMF, as shown in  Fig.\ref{fig-gauss-hmf-NFW-range2-simu}. In fact, varying the inner fitting range would also influence the fitting results. Considering the scale-dependent residuals between the best fitted NFW ESD and the stacked ESD in Fig\ref{fig-simu-prof}, this phenomenon is not surprising. 

During the fitting, to estimate the viral radius $R_{vir}$ for the $M_o$ formed mass bin, we would firstly repeat the single-NFW-fitting for several times. For each time, we fit the stacked ESD curve over a radius less than $R_{max}$. For the first time, $R_{max}$ is set to be 5 Mpc/h. After achieving the best fitted mass $M_{vir}$, we update the value of $R_{max}$ with the corresponding $R_{vir}$. After loop this operation for a few times, the value of $R_{vir}$ becomes stable.

In the above, we recover the underlying mean mass for the $M_o$ formed ranking bin with the stacked-NFW-fitting. However, the underlying mass recovered in this way differs from the $\overline{M}_t$, in which $M_t$ is set according to $M_{fof}$. So to make a more fair comparison, we choose to generate the reference RHMF also through fitting the stacked ESD profile, by modeling the halo mass distribution for the $M_t$ formed ranking bin. The resulting RHMF is shown as the red histogram in Fig.\ref{fig-gauss-hmf-NFW-range1-simu} and Fig.\ref{fig-gauss-hmf-NFW-range2-simu}. We point out that for other kind of mass definitions, such as $M_{200}$, $M_{vir}$ etc., almost the same stacked ESDs are got for the same ranking bin. Actually, we find that as long as the dispersion between these mass definitions are less than 0.15 dex, the shape and amplitude of their stacked ESD would be almost unchanged. Since these mass definitions are tight related,  the comparison between the reference RHMF and the reconstructed RHMF would not be affected by the choice of $M_t$.

\end{appendices}

\end{document}